\documentclass[prl,twocolumn,amsmath,amssymb,showpacs]{revtex4}
\usepackage{graphicx}
\usepackage{chngpage}

\begin{document}

\title{Bilayer graphene spectral function in RPA and self-consistent GW}
\author{A. Sabashvili, S. \"Ostlund, and M. Granath}
\affiliation{Department of Physics, University of Gothenburg,
SE-41296 Gothenburg, Sweden}

\date{\today}

\begin{abstract}
We calculate the single-particle spectral function for doped bilayer graphene in the low energy limit, described by two parabolic bands with zero band gap and long range Coulomb interaction. Calculations are done 
using thermal Green's functions in both the random phase approximation (RPA) and the fully self-consistent GW approximation. RPA (in line with previous studies) yields a spectral function which apart from the Landau quasiparticle peaks shows additional coherent features interpreted as plasmarons, i.e. composite electron-plasmon excitations. In GW the plasmaron becomes incoherent and peaks are replaced by much broader features. 
The deviation of the quasiparticle weight and mass renormalization from their non-interacting values is small which indicates that bilayer graphene is a weakly interacting system. The electron energy loss function, $Im[-\epsilon^{-1}_q(\omega)]$ shows a sharp plasmon mode in RPA which in GW approximation becomes less coherent and thus consistent with the weaker plasmaron features in the corresponding single-particle spectral function.
\end{abstract}

\pacs{} 

\maketitle

\section{I. Introduction}
\begin{figure}[b]
\includegraphics[scale=1]{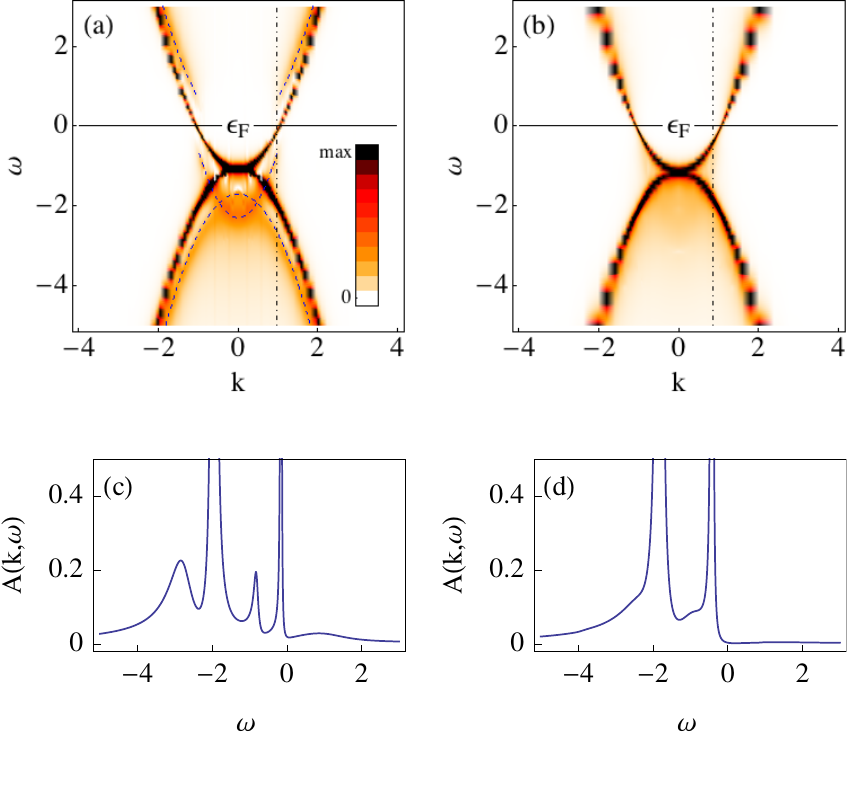}
\caption{(Color online) Single particle spectral function for bilayer graphene in the low energy limit at $r_s=3$. (a) - RPA, (b) - GW. The bare bands $\epsilon_{\vec k} = \pm k^2$ (in units $k_F = 1,\,  \epsilon_F = 1$) are rotationally symmetric (the patchy appearance is due to the finite k-space resolution). (c) and (d) are the cuts (dash-dotted lines) in (a) and (b), respectively. Dashed lines are guides to the eye for plasmaron dispersions.}
\label{fig:Grid8}
\end{figure}
Since its fabrication, graphene \cite{novoselov, dassarma, wallace, mcclure, neto} (a single layer of graphite) has been of interest for both theoreticians and experimentalists. It is a two-dimensional (2D) crystal with carbon atoms arranged on a honeycomb lattice with two sublattices. Due to its unique properties(e.g. high mobility even in highly doped 
cases) it opens new perspectives for engineering and is a candidate material for future nanoelectronic and spintronic devices \cite{geim}. 

The subject of this article is the closely related bilayer graphene, formed by stacking two graphene layers in Bernal $"AB"$ stacking  sequence in which the two layers are rotated by 60 degrees. These are coupled by interlayer tunneling between $A$ and $B$ sublattice sites with the hopping parameter $t_{\bot} \approx 0.39 \, eV $ \cite{kotov}. 

Bilayer graphene shares some features with both graphene and the ordinary two-dimensional electronic gas (2DEG). Its dispersion is quadratic, similar to a 2DEG but the effective Hamiltonian is chiral with zero band gap as in the case of graphene \cite{lv, kotov}. In both single layer and bilayer graphene the charge carrier density can be controlled by application of a gate voltage, a fundamental effect for potential technological applications \cite{ohta1, novoselov}.In addition, for bilayer graphene even the band gap is tunable with great potential for device applications \cite{ohta1, castro}.

Apart from the dispersion relation, the property which makes bilayer graphene different from that of a single layer is its coupling parameter being a function of the carrier density $r_s \sim n^{-1/2}$ \cite{sensarma, dassarma}. In other words the strength of Coulomb interaction is tunable, while the coupling parameter for the single layer graphene is constant $r_s \sim n^0$ and lies in the interval $0 \leq r_s \lesssim 2.2$. By comparing the values of $r_s$ for single- and bilayer graphene ($r_s \approx 68.5 \times 10^5/\sqrt{n}$, where $n$ is the number of carriers per $cm^{-2}$ with $n \approx 10^{9} - 5 \times 10^{12} \; cm^{-2}$) in vacuum it is clear that the strength of the Coulomb interaction can be much larger in bilayer graphene \cite{dassarma}.

The electronic structure of bilayer graphene \cite{johan, kotov} is characterised by the single particle spectral function $A_{\vec k}(\omega)$, which can be measured experimentally by angle resolved photo-emission spectroscopy (ARPES)\cite{ohta, bostwick}. It obeys the sum rule $\int \frac{d\omega}{2\pi} A_{\vec k}(\omega) = 1$ and can be interpreted as the probability distribution of an electron having momentum $\vec k$  and energy $\omega$. Sensarma \textit{et al}. \cite{sensarma} studied how Coulomb interaction affects the single particle spectral function of bilayer graphene away from half-filling. The authors used RPA to calculate that doped bilayer graphene is a Fermi liquid in the low energy limit, with a sharp quasiparticle peak. They also found additional weaker peak structures that they interpreted as plasmarons; a quasiparticle formed by the coupling between electron and plasmon, as originally predicted by B. Lundqvist 
\cite{lundqvist}.
Studying the physics of interaction between electrons and plasmons in graphene is particularly interesting because of recently proposed "plasmonic" devices that could merge photonics and electronics \cite{bostwick}.\\
\indent Experimentally plasmarons in the single layer graphene were observed by A. Bostwick {\em et al} \cite{bostwick} using angle-resolved photoemission spectroscopy. Apart from the two single particle crossing bands, two additional bands were observed and interpreted as a spectrum of plasmarons. The experimentally measured spectral function  compares qualitatively with that obtained within RPA.\\ 
\indent In this paper we compute numerically the single-particle spectral function $A_{\vec k}(\omega)$ for doped bilayer graphene in the low energy two-band approximation in both RPA and the fully self-consistent GW approximation \cite{hedin, hedin2, ferdi}. We use a thermal Green's function formalism, based on a finite set of imaginary frequencies and analytic continuation to real frequencies for the single-particle Green's functions in a controlled manner \cite{saba}.\\
\indent The results, presented in Fig. \ref{fig:Grid8}, show the spectral function with long lived Landau quasiparticles and satellite plasmaron peaks in RPA (Fig. \ref{fig:Grid8} (a)) and confirm the results of analytic calculations \cite{sensarma, sensarma1, kotov}, whereas in the GW approximation the plasmaron peaks are replaced by broad shoulders (Fig. \ref{fig:Grid8} (b)).
It has been emphatically argued that self-consistent GW underestimates the coherence of collective excitations \cite{barth, holm, ferdi, ayral} and our results showing a marked difference between the satellite peaks in RPA and GW most likely agree with this. Nevertheless we argue that the GW results are valuable  as a benchmark for more sophisticated self-consistent approaches including vertex corrections to the polarization.\\
\indent Below we describe our calculations in more detail.
\section{II. GW approximation}
\begin{figure}[h!]
\includegraphics[width=0.4\textwidth]{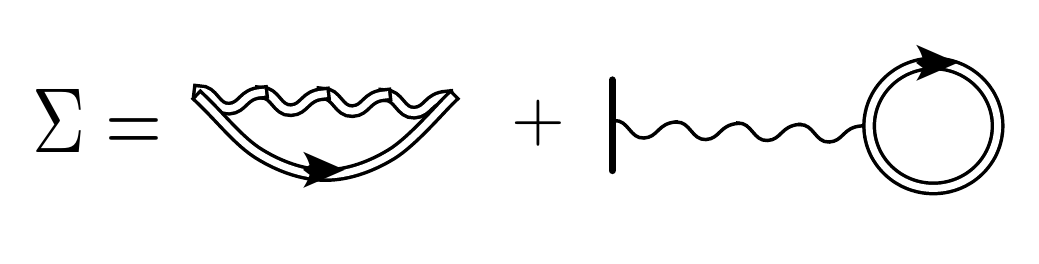}
\caption{Self-energy in the GW approximation. Double wiggly line, single wiggly line and double line correspond to dressed interaction, bare interaction and dressed Green's function, respectively}
\label{fig:SelfEnergy1}
\end{figure}

The GW approximation is derived perturbatively from the Hedin's equations \cite{hedin, hedin2}, giving the self-energy
\begin{equation}
\label{eq:GWapp}
\Sigma^{GW}_{\vec k}(i\omega_n) = \frac{1}{\beta}\int \frac{d^2q}{(2\pi)^2} \sum_{m=-\infty}^{\infty} W_{\vec q}(i\omega_m) G_{\vec k-\vec q}(i\omega_n- i\omega_m),
\end{equation}
where $W_{\vec q}(i\omega_n)$ and $G_{\vec k}(i\omega_n)$ are the dressed interaction and Green's function, respectively (all quantum numbers, such as momentum, spin, etc., are incorporated in $\vec k$ and $\vec q$). 
The argument of the Green's function is the fermionic Matsubara frequency $\omega_n = (2n + 1) \pi/\beta$, whereas the dressed interaction is a function of the bosonic Matsubara frequency $\omega_n = 2n \pi/\beta$ with $n$ integer. 
After computing $\Sigma^{GW}_{\vec k}(i\omega_n)$ (first diagram in Fig. \ref{fig:SelfEnergy1}) one should, in general, add the Hartree diagram (second diagram in Fig. \ref{fig:SelfEnergy1}) to it which in case of long-range Coulomb interaction gives zero contribution because it is cancelled by the positive background charge \cite{mattuck}.  
The approximation has the same form as the standard Hartree-Fock (HF) approximation. The difference is that the latter uses the bare Green's function and interaction, while the former is based on the dressed Green's function $G_{\vec k}(\omega)$ and dynamically screened (dressed) interaction. 

\begin{figure}[h!]
\includegraphics[width=0.45\textwidth]{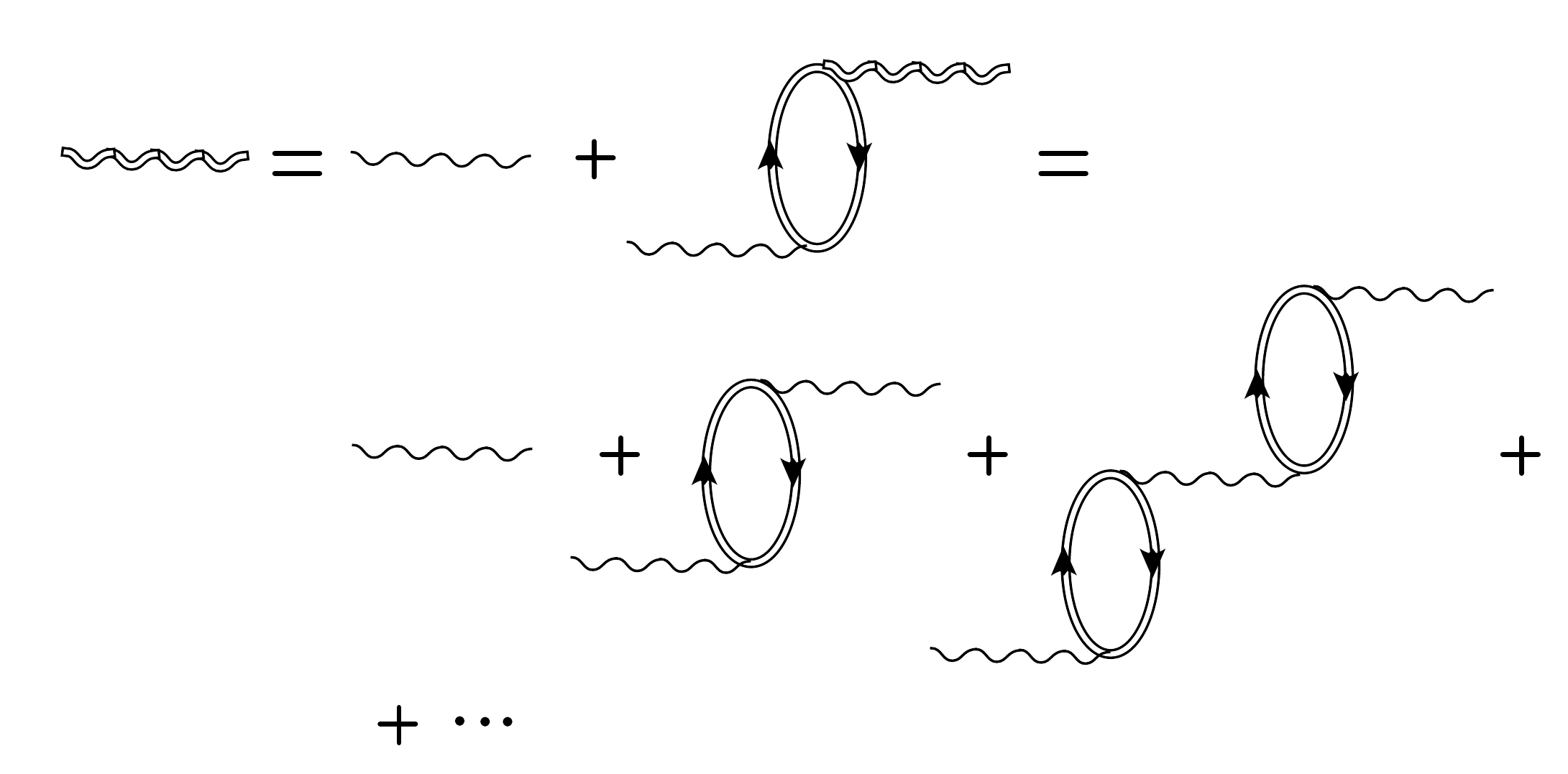}
\caption{Screened interaction in the GW approximation is given by geometric series. Bubble diagram represents polarization $\Pi_{\vec q}(i\omega_n)$.}
\label{fig:EffInteractionDiagram}
\end{figure}

The screened interaction $W_{\vec q}(i\omega_n) $ is an infinite geometric series of diagrams  (Fig. \ref{fig:EffInteractionDiagram}) consisting of the bare interaction $V_q$ and the irreducible polarization diagram $\Pi_{\vec q}(i\omega_n)$ which in GW is given by 
\begin{equation}
\label{eq:pi}
\Pi_{\vec q}(i\omega_n) = -g \int \frac{d^2k}{(2\pi)^2} \frac{1}{\beta}\sum_{m=-\infty}^{\infty} G_{\vec k}(i\omega_m) G_{\vec k+\vec q}(i\omega_n + i\omega_m),
\end{equation}
where $g$ is the degeneracy factor. In RPA it is computed using the same relation but with the full Green's functions replaced by the bare ones. So, the RPA polarization is just the zeroth order term in the expansion of $\Pi_{\vec q}(\omega_n)$ in the bare interaction.
After summing up the geometric series one obtains the following expression for the screened interaction
\begin{equation}
\label{eq:effinteraction}
W_{\vec q}(i\omega_n) = \frac{V_q}{1 + V_q\Pi_{\vec q}(i\omega_n)}.
\end{equation}



The effective bare Coulomb interaction for bilayer graphene is given by $V_q = \frac{2 \pi e^2}{\kappa q}$  
where $\kappa$ represents the background dielectric constant.\cite{johan} Using $E_F$ and $k_F$ as units of energy and momentum, respectively enables us to write $V_q$ in terms of the dimensionless coupling parameter $r_s= e^2 g m/(k_F \kappa)$:
\begin{equation}
V_q = \frac{\pi r_s}{q}.
\end{equation}

\section{III. Bilayer graphene}
\subsection{A. Effective model}
The low energy limit of bilayer graphene is valid if the scale of all relevant energies are smaller than the interlayer hopping parameter $t_{\bot}$ such that the two outer bands can be ignored. Incorporating the two layers as an additional index, the Hamiltonian can be represented as a 2$\times$2 matrix and thus four-band model is reduced to the effective two-band model with the total degeneracy of $g=4$ (due to the spin and valley index) \cite{falko, kotov},
\begin{equation}
H_0 = - \frac{1}{2m} \left( \begin{array}{cc} 
                                            0 & (k_x + ik_y)^2\\
                                             (k_x - ik_y)^2 & 0
                                             \end{array} \right).
\end{equation}
It is clear that the corresponding energy spectrum is parabolic,
\begin{equation}
\epsilon_k = \pm \frac{k^2}{2m}.
\end{equation}
$m=t_{\bot}/(2v^2_F)\approx0.054 m_e$ is the effective mass of the electron in the low energy limit, $m_e$ being the free electron mass. Corresponding free Fermionic Matsubara Green's function is
\begin{equation}
\hat{G}^0_{\vec{k}}( i\omega_n) = (i\omega_n - H_0 + \mu)^{-1} = \frac{1}{2}\sum_{s=\pm} \frac{\mathbb{1} + s \hat{\sigma}_{\vec{k}}}{i\omega_n - s |\epsilon_k| + \mu},
\end{equation} 
$s$ indexes the conductance and valence bands, $\mu$ is the chemical potential and $\sigma_{\vec{k}}$ is given by
\begin{equation}
\hat{\sigma}_{\vec{k}} = \sum_{j = \pm} \frac{k^2_{j}}{k^2} \hat{\sigma}_j = \sum_{j = \pm} e^{j2\theta_{\vec{k}}} \hat{\sigma}_j,
\end{equation}
where $k_{\pm} = k_x \pm ik_y$, $\hat{\sigma}_{\pm} = (\hat{\sigma}_1 \pm i\hat{\sigma}_2)/2$, $\hat{\sigma}_1$ and $\hat{\sigma}_2$ are Pauli matrices, $\theta_{\vec{k}}$ is the angle of the vector $\vec{k}$ with respect to the x-axis. 

Let us rewrite Eq. \ref{eq:pi} for the two-band model. After summing over all internal indicies (which in this case is just the matrix index of $\hat{G}_{\vec{k}}( i\omega_n)$ ) when computing the polarization diagram we obtain the following expression for $\Pi_{\vec{q}}(i\omega_n)$,
\begin{align}
\label{eq:PiTB}
\Pi_{\vec{q}}(i\omega_n) =  -\frac{g}{\beta} \int \frac{d^2k}{(2\pi)^2}\sum_{m=-\infty}^{\infty} Tr (\hat{G}_{\vec{k}}(i\omega_m) \times \nonumber
 \\ \times \hat{G}_{\vec{k}+\vec{q}}(i\omega_n + i\omega_m)).
\end{align}
In the non-interacting limit $\Pi_{\vec{q}}(i\omega_n)$ can be written in the following simple form:
\begin{equation}
\Pi^0_{\vec{q}}(i\omega_n) =  - g \sum_{s,s'}\int \frac{d^2k}{(2\pi)^2} \frac{(f^s_{k} - f^{s'}_{k+q}) F_{s,s'}(\vec{k},\vec{k}+\vec{q}) }{i\omega_n + |\epsilon_k| - |\epsilon_{k+q}| + \mu}, 
\end{equation}
where $f^s_{k} = 1/(1 + e^{\beta(\epsilon_k - \mu)})$ is the Fermi distribution function and 
\begin{align*}
F_{s,s'}(\vec{k},\vec{k}+\vec{q}) & = \frac{1}{4} Tr(1 + s\hat{\sigma}_{\vec{k}})(1 + s'\hat{\sigma}_{\vec{k}+\vec{q}}) \\[2mm]
       & = \frac{1}{2}(1 + ss'\cos(2\theta_{\vec{k},\vec{k}+\vec{q}})),
\end{align*}
with $\theta_{\vec{k},\vec{k}+\vec{q}}$ being the angle between the vectors $\vec{k}$ and $\vec{k}+\vec{q}$.  $\Pi_{\vec{q}}(i\omega_n)$ is an angle-independent function which can be seen by extracting the angle $\theta_{\vec{q}}$ using the rotation of the integration variable in Eq. \ref{eq:PiTB} with $\theta_{\vec{q}}$. Consequently, the screened interaction is angle-independent as well.
Note that since the polarization is a scalar due to the trace in Eq. \ref{eq:PiTB} the screened interaction remains to be a scalar quantity as well.

Clearly, the GW self-energy is generalised to
\begin{equation}
\label{eq:SEtb}
\hat{\Sigma}^{GW}_{\vec{k}}(i\omega_n) = \frac{1}{\beta}\int \frac{d^2q}{(2\pi)^2} \sum_{m=-\infty}^{\infty} W_q(i\omega_m) \hat{G}_{\vec{k}-\vec{q}}(i\omega_n- i\omega_m)
\end{equation}
In order to the see the matrix structure of the $\hat{\Sigma}^{GW}_{\vec{k}}(i\omega_n)$ we perform the following integration variable transformations in Eq. \ref{eq:SEtb} with $\hat{G}_{\vec{k}-\vec{q}}$ replaced by  $\hat{G}^0_{\vec{k}-\vec{q}}$,
\begin{align}
&\vec{q}_1= \vec{k} - \vec{q}
\end{align}
and
\begin{align}
& \vec{q}_2= \mathcal{R}(\pi + \theta_{\vec{k}}) \vec{q}_1,
\end{align}
where $\mathcal{R}(\pi + \theta_{\vec{k}}) $ denotes the rotation matrix with the angle $\pi + \theta_{\vec{k}}$. Therefore the self-energy can be rewritten as
\begin{equation}
\hat{\Sigma}^{GW}_{\vec{k}}(i\omega_n)  = \frac{1}{2} \left( \begin{array}{cc} 
                                          \Sigma_0 & \Sigma_+ e^{i 2 \theta_{\vec{k}} } \\
                                           \Sigma_- e^{- i 2 \theta_{\vec{k}} } & \Sigma_0
                                             \end{array} \right),
\end{equation}
with
\begin{align}
&\Sigma_{0} =  \int \frac{d^2q}{(2\pi)^2} a^+_{q_2} W(|\vec{k} - \vec{q_2}|)
\end{align}
and
\begin{align}
& \Sigma_{\pm} = \int \frac{d^2q}{(2\pi)^2} a^-_{q_2} W(|\vec{k} - \vec{q_2}|) e^{\pm i 2 \theta_{\vec{q}_2}  }.
\end{align}
Now, if one makes the variable transformation $\theta_{\vec{q}_2} = -\theta_{\vec{q}_2}$ in $\Sigma_+$ ($\Sigma_-$) it becomes obvious that $\Sigma_+ = \Sigma_-$ which means that $\hat{\Sigma}^{GW}_{\vec{k}}(i\omega_n)$ and the fully interacting Green's function $\hat{G}_{\vec{k}}(i\omega_n)$ have and retain the same structure as the one of the free Green's function $\hat{G}^0_{\vec{k}}(i\omega_n)$ (Eq. \ref{eq:G0matrix}, \ref{eq:G0matrix2}) throughout the whole self-consistent calculation. 
\begin{align}
\label{eq:G0matrix}
&\hat{G}^0_{\vec{k}}( i\omega_n) = \frac{1}{2} \left( \begin{array}{cc} 
                                            a^+_k & a^-_k e^{i2\theta_{\vec{k}} }\\
                                            a^-_k e^{-i2\theta_{\vec{k}} } & a^+_k
                                             \end{array} \right) \\[2mm]
\label{eq:G0matrix2}
&a^{\pm}_k \equiv (i\omega_n - |\epsilon_k| + \mu)^{-1} \pm (i\omega_n + |\epsilon_k| + \mu)^{-1} 
\end{align}
So, it is sufficient to set up calculations for $\hat{\Sigma}^{GW}_{\vec{k}}(i\omega_n)$ and $\hat{G}_{\vec{k}}(i\omega_n)$ only at $\theta_{\vec{k}} = 0$.

We start the self-consistent calculation by discretising momenta and angles. Since our interest is focused on the low energy properties the absolute value of $\vec k$ ranging from 0 to 4 is discretised into 40 points logarithmically giving denser number of points around $k_F$. The rest of the integration variables ($|\vec q|$, $\theta_{\vec k}$ and $\theta_{\vec q}$) are discretized linearly. $|\vec q|$ is discretized into 80 points and lies in the interval $[1/80, \, 4]$ while the number of the discretization points for $\theta_{\vec k}$ and $\theta_{\vec q}$ is 10. First, the free Greens's function is evaluated at $\theta_{\vec k}=0$ and then it is rotated by an angle $\theta_{\vec k}$ in order to obtain the polarization (Eq. \ref{eq:PiTB}). Then the screened interaction is computed using Eq. \ref{eq:effinteraction} which enables us to evaluate the GW self-energy (Eq. \ref{eq:SEtb}). After calculating $\hat{\Sigma}^{GW}_{\vec{k}}(i\omega_n) $ at  $\theta_{\vec k}=0$ we update the Green's function through Eq. \ref{eq:Gptb}. This is done repeatedly: if the procedure converges to a fixed point, a solution has been found. 
The calculations are done at $T/\epsilon_F=1/10$ with $N=121$ number of Matsubara frequencies.
\subsection{B. Periodized Green's functions}
The GW approximation implies a self-consistent numerical calculation, which may be solved iteratively \cite{strand}.
Obviously these calculations include very demanding operations including infinite sums over Matsubara frequencies. In order to cope in numerical calculations with these kinds of problems we use a new formalism for finite temperature fermionic thermal Green's functions in the single band case described in \cite{saba} and summarised below. 

Performing numerical calculations using thermal Green's functions \cite{abrikosov, N_O} may be done by the discretization of imaginary time. Since the fermionic thermal Green's function is anti-periodic over  $\tau \in [-\beta, \beta]$ domain with the period $\beta$ we discretize the interval $\tau \in [0, \beta]$ into $N$ evenly spaced points, $\tau = (\beta/N) j, \; j = 1,..., N-1$.
Due to the discontinuity of the fermionic Green's function at $\tau = 0$ (limits $\tau \rightarrow 0^-$ and $\tau \rightarrow 0^+$ differ from each other) some specific value must be assigned to $G_k(\tau_j = 0)$ when doing numerical computations. We define $G_k(\tau_j = 0)$ by the average of $G_k(\tau = 0^-)$ and $G_k(\tau = 0^+)$.
After applying discrete Fourier transformation to the non-interacting thermal Green's function
\begin{equation}
G^0_k(\tau) = e^{-\epsilon_k \tau}[(n_k - 1) \theta (\tau) + n_k \theta(-\tau)]
\end{equation}
where $n_k = \langle c^{\dagger}_k \, c_k \rangle$ is the occupation number we obtain periodic set of the Green's function values in the Matsubara frequency space 
\begin{align}
\label{eq:discretizedGF}
G^0_k(i\omega_n)  = \eta  \coth  \eta  (i\omega_n - \epsilon_k).
\end{align}
Here $\eta \equiv \frac{\beta}{2N}$ and $\epsilon_k$ is a single-particle excitation spectrum for a given model. It is obvious that Eq. \ref{eq:discretizedGF} is periodic under  $i\omega_n \rightarrow i\omega_n + i\Omega_N, \; \Omega_N \equiv \frac{\pi}{\eta}$.
The periodized full Green's function is given by
\begin{align}
\label{eq:fullperiodicGF}
 G_k(i\omega_n)  = \eta  \coth  \eta (i \omega_n - \epsilon_k - \Sigma_k(i\omega_n)),
\end{align}
which has the correct non-interacting limit and together with $G^0_k(i\omega_n)$ yields standard continuum expression for the Greens function as $N$ tends to $\infty$ ( $\eta \rightarrow 0$ ).  Due to the nontrivial hyperbolic function in Eq. \ref{eq:fullperiodicGF} one can not define the self-energy using simply $G_0^{-1}$ and $G^{-1}$ as it is done in the standard theory. In this case the self-energy is defined by the amputated skeleton diagrams (\cite{N_O}, see Sec. 5.1) and through Eq. \ref{eq:fullperiodicGF}.

In the case of two band model $G_k(i\omega_n)$ is generalised to
\begin{equation}
\label{eq:Gptb}
\hat{G}_{\vec{k}}(i\omega_n) = \eta \coth\eta( (\hat{G}^0_{\vec{k}}( i\omega_n))^{-1} - \hat{\Sigma}_{\vec{k}}(i\omega_n)),
\end{equation}
where
\begin{align*}
(\hat{G}^0_{\vec{k}}( i\omega_n))^{-1} & = \left(\begin{array}{cc}
                                                         i\omega_n + \mu & - |\epsilon_k|\, e^{i2\theta_{\vec{k}}}\\
                                                         -|\epsilon_k| \, e^{-i2\theta_{\vec{k}}} &  i\omega_n + \mu 
                                                   \end{array} \right) \nonumber \\[2mm]
              & = (i\omega_n + \mu) \mathbb{1} - |\epsilon_k| \hat{\sigma}_{\vec{k}}.
\end{align*}
The periodized Green's function for both single and two-band cases is consistent with the corresponding Luttinger-Ward $\Gamma$-functional \cite{ward, baym} (the former is consistent with the $\Gamma$-functional as presented in Eq. 4 in \cite{saba} while the former - with the same equation where $G_k(i\omega_n)$ is replaced by $\hat{G}_{\vec{k}}(i\omega_n)$). 

To perform analytic continuation for $G_k(i\omega_n)$ we first rewrite it by means a conformal transformation in a new basis where it can be represented as a sum of simple poles. Then the Pad\'e method \cite{vidberg} of fitting to a rational function is used which enables us to evaluate the Green's function on the real frequency axis. In the case of a two-band model the trace of $\hat{G}_{\vec{k}}( i\omega_n)$ is used as an input to the same procedure of analytic continuation as the one carried out for $G_k(i\omega_n)$. 

\subsection{C. Spectral function}
\begin{figure}[h!]
\includegraphics[scale=1]{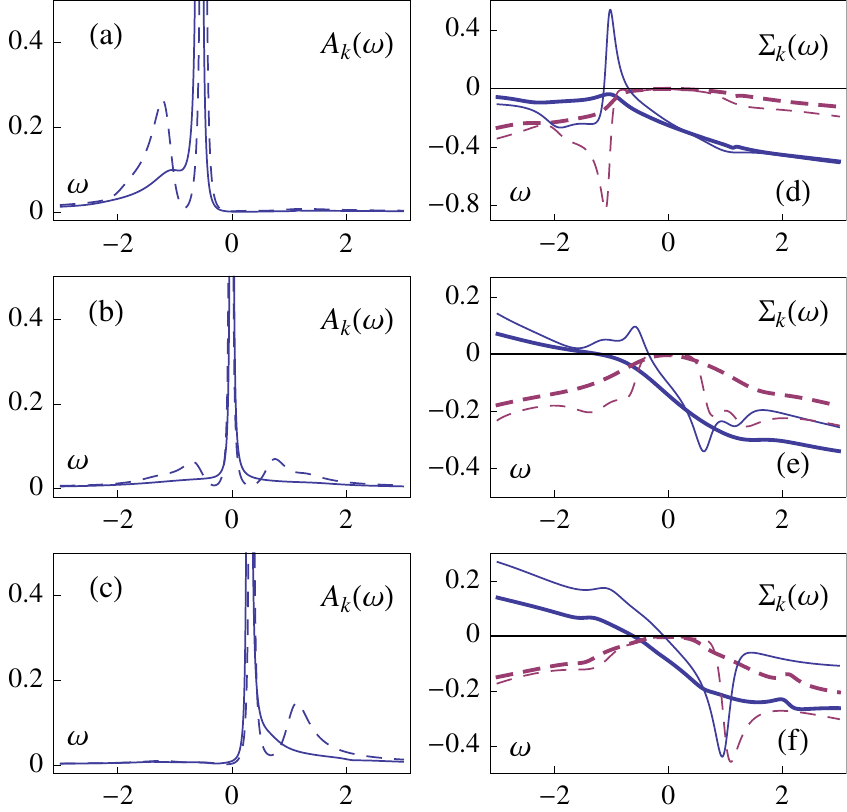}
\caption{(Color online) $r_s=3$. Left column: spectral weight in RPA (dashed line) and GW approximation (solid line) at $k \approx 0.76 k_F$ (a), $k=k_F$ (b), $k \approx 1.20k_F$ (c). Right column: the real (blue solid line) and imaginary (red dashed line) part of the self-energy at $k \approx 0.76 k_F$ (d), $k=k_F$ (e), $k \approx 1.20k_F$ (f) in RPA (thin line) and GW approximation (thick line).}
\label{fig:Grid4}
\end{figure}

\begin{figure}[h!]
\includegraphics[scale=1]{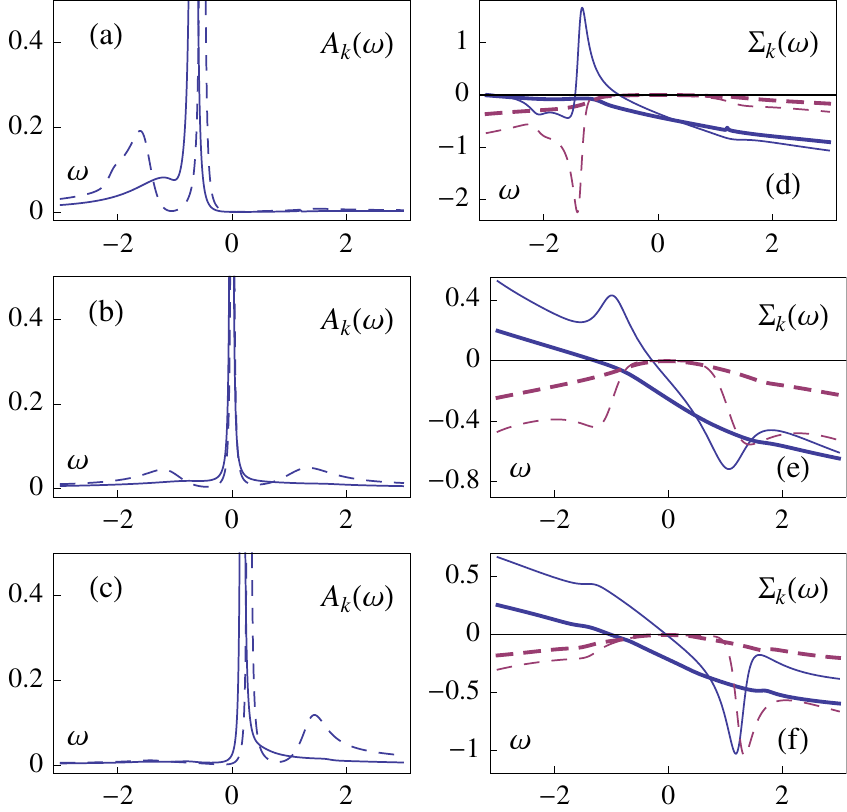}
\caption{(Color online) $r_s=7$. Left column: spectral weight in RPA (dashed line) and GW approximation (solid line) at $k \approx 0.76 k_F$ (a), $k=k_F$ (b), $k \approx 1.20k_F$ (c). Right column: the real (blue solid line) and imaginary (red dashed line) part of the self-energy at $k \approx 0.76 k_F$ (d), $k=k_F$ (e), $k \approx 1.20k_F$ (f) in RPA (thin line) and GW approximation (thick line).}
\label{fig:figure3}
\end{figure}

The spectral function is given by
\begin{align}
&A_{\vec{k}}(\omega) = - \frac{1}{\pi} Im[ Tr \hat{G}_{\vec{k}}(\omega + i 0^+)]. 
\end{align}
where we perform analytic continuation after applying trace to $\hat{G}_{\vec{k}}(i\omega_n)$.
In order to study the low energy properties we also compute the spectral function projected on the conductance band 
\begin{align}
A_{\vec{k}}(s= +, \omega) = - \frac{1}{\pi}  Im [G_{\vec{k}}(s = +, \omega + i0^+)],
\end{align}
where $G_{\vec{k}}(s = +, \omega)$ represents the eigenvalue of $\hat{G}_{\vec{k}}(i\omega_n)$ corresponding to the upper band after analytic continuation to the real axis.
In Fig. \ref{fig:Grid4} and \ref{fig:figure3} we present the spectral functions (left column) for different values of $k_F$ together with the corresponding self-energies (right column) in the GW approximation and RPA at $r_s = 3$ and $r_s=7$, respectively. 

As the plots  show the spectral weight in the RPA away from $k_F$ has two peaks: the main Landau quasiparticle peak and plasmaron peaks. 
The presence of the plasmaron excitation also give jumps in the real and imaginary parts of the corresponding self-energies. The RPA plasmaron excitation has lower weight at $r_s=7$ than the one at $r_s=3$, although the spectral functions have qualitatively same behaviour which is also noticeable in the case of the GW approximation.  Most of the structure obtained in the RPA is not presented in the GW approximation. We interpret this as being due to stronger screening in GW.

\begin{figure}[h!]
\includegraphics[scale=1]{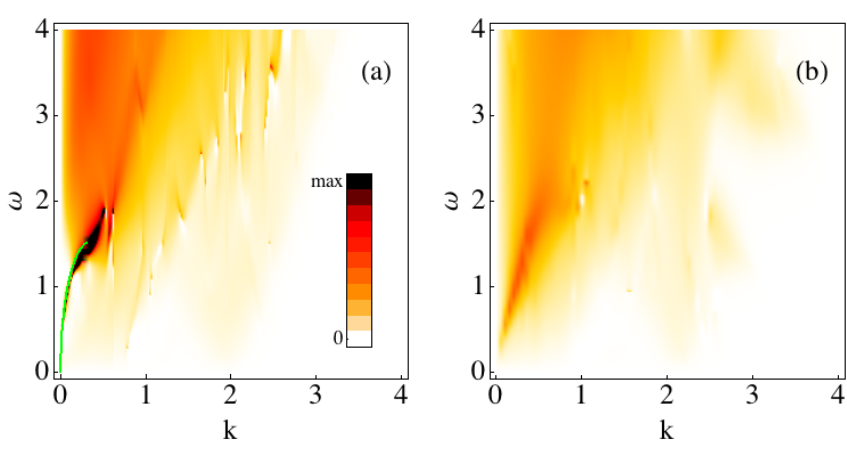}
\caption{(Color online) $Im[-\epsilon^{-1}_q(\omega)]$ in RPA (a) and GW approximation (b) at $r_s=7$ (same color intensity scale on both plots). Green solid line in (a) represents the plasmon dispersion expanded up to the second order in $q$. The unexpected discontinuities are artificial and due to difficulties with the analytic continuation of a two-particle function.}
\label{fig:figure5}
\end{figure}

In Fig. \ref{fig:figure5}(a) the electron energy loss spectrum $Im[-\epsilon^{-1}_q(\omega)]$ ($\epsilon_q(\omega) = 1 + V_q \Pi_{\vec q}(\omega)$ - dielectric function) in RPA is plotted showing the plasmaron dispersion relation (black color) which is in a quite good agreement for small $q$-values with its analytic version (solid line) expanded up to second order in $q$ \cite{dassarma, sensarma1},
\begin{equation*}
\omega_q \simeq e \sqrt{\frac{gE_Fq}{\kappa}}\left(1 - \frac{r_sq}{8k_f}\right).
\end{equation*}
$Im[-\epsilon^{-1}_q(\omega)]$ was also calculated in the GW approximation (Fig. \ref{fig:figure5}(b)) 
where the plasmon mode is less coherent than that in  RPA which is in agreement with the fact that the plasmaron features in the GW spectral function are weaker than in RPA.
\subsection{D. Quasiparticle weight and effective mass}
The quasiparticle weight $Z$ and renormalized mas $m^*$, given in Table I, are computed for both the GW and RPA approximations using the formulas:
\begin{align}
\label{eq:qp}
& Z = \frac{1}{1 - \frac{\partial Re\Sigma_{k_F}(\omega)}{\partial \omega}|_{\epsilon_F}},\\[2mm]
\label{eq:EffMass}
& \frac{m^*}{m} = \frac{Z^{-1}} {1 + \frac{m}{k_F}  \frac{\partial Re\Sigma_{k}(\omega = \epsilon_F)}{\partial k}|_{k_F} }.
\end{align}
As expected, the quasiparticle weight decreases with increasing interaction strength because the interaction shifts the weight from the coherent quasiparticle peak through incoherent scattering. Since the GW approximation does not yield the plasmaron peaks and the interaction gets more screened, most of the weight is concentrated in the Landau quasiparticle which results in a bigger quasiparticle weight than that in the case of RPA. The mass renormalization is less than $7\%$ in both approximations meaning that we are dealing with a weakly interacting system. 
\begin{table}[h!]
\begin{minipage}[t]{0.3\linewidth}
\begin{adjustwidth}{-1.3cm}{}
\begin{tabular}{ l c c  }
     & $Z $ & $m^*/m$  \\
    \hline
    RPA & \, 0.798 & \, 0.978 \\ 
    GW & \, 0.851 & \,  0.946 
    \end{tabular}
\end{adjustwidth}
\end{minipage}
\hspace{0.5cm}
\begin{minipage}[t]{0.3\linewidth}
\begin{adjustwidth}{-0.4cm}{}
\begin{tabular}{ l c c  }
     & $Z $ & $m^*/m$  \\
    \hline
    RPA &\, 0.685 & \, 0.986 \\ 
    GW & \, 0.806 & \, 0.929
    \end{tabular}
\end{adjustwidth}
\end{minipage}
\caption{Quasiparticle weight $Z$ and effective mass relative to the one of the free electron $m^*/m$ at $r_s=3$ (left) and $r_s=7$ (right).} 
\end{table}
By comparing our results with the ones presented in \cite{sensarma} one can see that the agreement is quite good. 

\section{IV. Conclusion}
We present the single particle spectral function and self-energy for bilayer graphene in the low energy limit as described with a two band model. Calculations are done in both RPA and self-consistent GW using a discretized thermal Green's function formalism. 
In RPA, the spectral function and energy loss spectrum show prominent plasmaron peaks and sharp plasmon mode, respectively whereas in GW the plasmaron peaks are replaced by broad shoulders which is consistent with a less coherent plasmon mode. The RPA spectral function, quasiparticle weight and effective mass are in a good agreement with those in \cite{sensarma} computed using a conventional Matsubara Green's function method.
\section{Acknowledgements}
We acknowledge useful discussions with R. Sensarma. This work was supported by $MP^2$ platform at the University of Gothenburg and Swedish Research Council (grant no. 2011-4054 and 2008-4242).

\end{document}